\documentclass{optica-article}

\journal{oe}

\articletype{Research Article}

\usepackage{bm}

\begin{document}

\title{Spatial-photonic Ising machine by space-division multiplexing with physically tunable coefficients of a multi-component model}

\author{Takumi Sakabe\authormark{1,*}, Suguru Shimomura\authormark{1}, Yusuke Ogura\authormark{1}, Ken-ichi Okubo\authormark{2}, Hiroshi Yamashita\authormark{1}, Hideyuki Suzuki\authormark{1}, and Jun Tanida\authormark{1}}

\address{\authormark{1}Graduate School of Information Science and Technology, Osaka University, 1-5 Yamadaoka, Suita, Osaka 565-0871, Japan}
\address{\authormark{2}Department of Electrical Engineering, Sanyo-Onoda City University, 1-1-1 Daigakudori, Sanyo-Onoda, Yamaguchi 756-0884, Japan}

\email{\authormark{*}t-sakabe@ist.osaka-u.ac.jp} 



\begin{abstract*}
This paper proposes a space-division multiplexed spatial-photonic Ising machine (SDM-SPIM) that physically calculates the weighted sum of the Ising Hamiltonians for individual components in a multi-component model.
Space-division multiplexing enables tuning a set of weight coefficients as an optical parameter and obtaining the desired Ising Hamiltonian at a time. 
We solved knapsack problems to verify the system's validity, demonstrating that optical parameters impact the search property.
We also investigated a new dynamic coefficient search algorithm to enhance search performance.
The SDM-SPIM would physically calculate the Hamiltonian and a part of the optimization with an electronics process. 
\end{abstract*}

\section{Introduction}
Combinatorial optimization problems exist across diverse fields in science, industry, and society \cite{A.Lucas}. 
However, these are often NP-hard problems, and the computational cost to find the optimal solution grows exponentially with the problem’s size. 
Various algorithms have been developed to find optimal solutions, such as simulated annealing (SA) \cite{S.Kirkpatrick} and evolutionary computation \cite{H.Muhlenbein}.
Subsequently, equations based on the Ising model were proposed to solve combinatorial optimization problems efficiently.
Ising machines are dedicated hardware devices that use physical pseudo spins to implement the Ising model and minimize the energy in the model \cite{N.Mohseni}. 
The Hamiltonian of the Ising model without an external magnetic field is represented by
\begin{equation}
\label{eqI}
\mathcal{H}=-\sum^N_{j\ h} J_{jh}\sigma_j \sigma_h,
\end{equation}
where $\sigma_j \in \{-1,1\} (j=1,2,\cdots,N)$ are spin variables and $J_{jh}(j,h=1,2,\cdots,N)$ are spin-spin interactions. 
$N$ is the number of spins.

Ising machines are developed using various physical systems, which can solve combinatorial optimization problems quickly. 	
Because the ability to represent the problem in an Ising model depends on the flexibility of the spin-spin interaction $J$, using a suitable physical system is vital for the developing practical machines.
An example of a phenomenon used in such systems is the quantum mechanical effect, associated with superconducting circuits \cite{M.W.Johnson} and trapped ions \cite{K.Kim}. 
These methods use quantum annealing \cite{T.Kadowaki} based on quantum fluctuation. 
SA has also been emulated in semiconductor integrated circuits, including CMOS annealing machines \cite{M.Yamaoka} and digital annealers \cite{M.Aramon}. 
Among implementations using physical phenomena, photonics-based implementation approaches are considered one of the most promising for addressing large-scale problems, because of the capabilities of light in parallel and high-speed processing and the robustness of computing systems \cite{C.Li}. 
One example is a coherent Ising machine \cite{T.Inagaki,T.Honjo}, where pseudo spins are implemented with optical pulses generated by degenerate optical parametric oscillators \cite{A.Marandi}.
Another is an integrated nanophotonic recurrent Ising sampler \cite{M.Prabhu}, where pseudo spins are implemented with coherent optical amplitudes.

A promising technique for controlling light on a large scale is spatial light modulation, (often used for computing), which exploits the parallel propagation characteristics of light \cite{X.Lin,J.Chang,J.Bueno}. 
A promising use of this optical strategy is the spatial-photonic Ising machine (SPIM) \cite{D.Pierangeli:19}, which represents spins through the modulation of light waves using a spatial light modulator (SLM). 
Spin-spin interactions are achieved by superimposing light waves through free-space propagation. 
Compared with other physical implementation systems, SPIMs provide a simpler configuration and high scalability in handling the number of spins, because they use the propagation parallelism of light based on Fourier optics \cite{J.W.Goodman}. 
Because of these features, SPIMs have attracted much attention, and many systems and methods to enhance the SPIM have been considered.
For example, annealing methods \cite{D.Pierangeli:20a,D.Pierangeli:20b}, a spin encoding method \cite{J.Huang}, an interaction model using the transmission matrix of the scattering medium \cite{G.Jacucci}, and that using nonlinear optical effects \cite{S.Kumar} have been proposed. 

However, the SPIM scheme has limited flexibility in expressing interaction $J$ to solve combinatorial optimization problems; its primitive version can only solve problems with a rank-1 interaction matrix. 
This is a limitation for practical applications because only a few types of problems can be mapped.
Several methods have been developed to remove this limitation, including (1) the quadrature photonic spatial Ising machine that introduces orthogonal phase modulation and external magnetic fields \cite{W.Sun}, (2) the phase-encoding and intensity detection Ising annealer, which uses computer holograms to perform calculations for each spin \cite{J.Ouyang}, and (3) a new computing model using the Gauge transformations implemented by wavelength multiplexing \cite{L.Luo}.
These methods have advantages for the representation of problems but their scalability in handling spins is reduced.
We address this issue by proposing, a multi-component model of SPIM that uses time-division multiplexing \cite{H.Yamashita}. 
In the multi-component model, the rank of the interaction matrix to be implemented is improved by decomposing the Ising Hamiltonian into several components that can be handled in the primitive SPIM scheme.
For general combinatorial optimization problems, the coefficient weights of components are important because they affect the quality of the obtained solution and convergence speed in executing the Ising machine.
For controlling the coefficients, each component of the Ising Hamiltonian is obtained by time-division multiplexing.
Then the individual components multiplied with each coefficient weight are summed on a computer.
Consequently, the overall Ising Hamiltonian could not be obtained in a batch, and all components must be acquired separately.

This paper presents a space-division multiplexed SPIM (SDM-SPIM), as an option of the system configurations to optically calculate a sum of multi-component Hamiltonian at a time while maintaining the high flexibility of the interaction matrix.
In the space-division multiplexing scheme, the beams encoding each component are independently controlled to adjust the individual optical intensities, and thus the weight coefficients can be physically multiplied simultaneously.
Then, the sum of the Ising Hamiltonian for each component can be obtained, by superimposing these beams. 
Furthermore, SDM-SPIM makes it possible to physically tune an optical parameter, a set of weight coefficients relating to constraint conditions of problems, and a part of the optimization process can be replaced by a physical process that dynamically tunes an optical parameter.
This study aims to validate the method and its capability using physical parameter tuning, which is achieved by implementing an SPIM with spatial-division multiplexing. 
We constructed a proof-of-concept system and applied it to knapsack problems---combinatorial optimization problems with a constraint term.
Furthermore, we analyzed the impact of physical parameters on the search characteristics of our method and investigated methods to enhance the search performance in the SDM-SPIM.

\section{Method}
\subsection{SPIM with space division multiplexing}
In SPIMs, the Ising Hamiltonian is optically calculated by phase modulation of an amplitude-distributed light. 
The obtained Ising Hamiltonian is evaluated and used to update the next spin state.
The spin $\sigma_j$ is encoded into phase modulation $\phi_j \in \{0,\pi\}$ by an SLM, where $\sigma_j = \exp(\imath\phi_j)=\pm1$.
When the amplitude distribution of light is $\xi_j$, the element $J_{j h}$ of the interaction matrix $J$ is proportional to $\xi_j$ and $\xi_h$. 
The Ising Hamiltonian in SPIMs can be expressed as follows \cite{D.Pierangeli:19}.
\begin{equation}
\label{eqH}
\mathcal{H} \propto \sum^N_{j\ h}\xi_j\xi_h\sigma_j\sigma_h.
\end{equation}
The Ising model with such a Hamiltonian is known as the Mattis model \cite{D.C.Mattis}.
The interaction matrix $J$ is limited to the form $J\propto \bm{\xi} \bm{\xi}^T $, where $\bm{\xi}=(\xi_1,\cdots,\xi_N)^T$, and SPIMs can accommodate only a real symmetric matrix of rank 1 as the interaction matrix. 

This study adopts a multi-component model based on the linear combination of the Mattis model to represent arbitrary Hamiltonians \cite{H.Yamashita}. 
The Ising Hamiltonian is represented by
\begin{equation}
\label{eqHM}
\mathcal{H}=-\sum_{k}^K \alpha^{(k)} \sum^N_{j\ h} \xi^{(k)}_{j}\xi^{(k)}_{h} \sigma_j \sigma_h,
\end{equation}
where $K$ is the total number of multiplexes, $\alpha^{(k)}$ are weight coefficients of $k$-th terms, and $\xi^{(k)}_{j}$ are the amplitude distributions corresponding to the multiplexing number. 
An Ising model with an interaction matrix of rank $K$ can be represented by this equation. 
In Eq. \eqref{eqHM}, the spin variable $\sigma_j$ is common to the multiplexed terms, and the SLM to manipulate spins can be shared by the multiplexed lights. 
In contrast, the amplitude distribution $\xi^{(k)}_{j}$ should be treated independently.

\begin{figure}[t]
\centering\includegraphics[width=11cm]{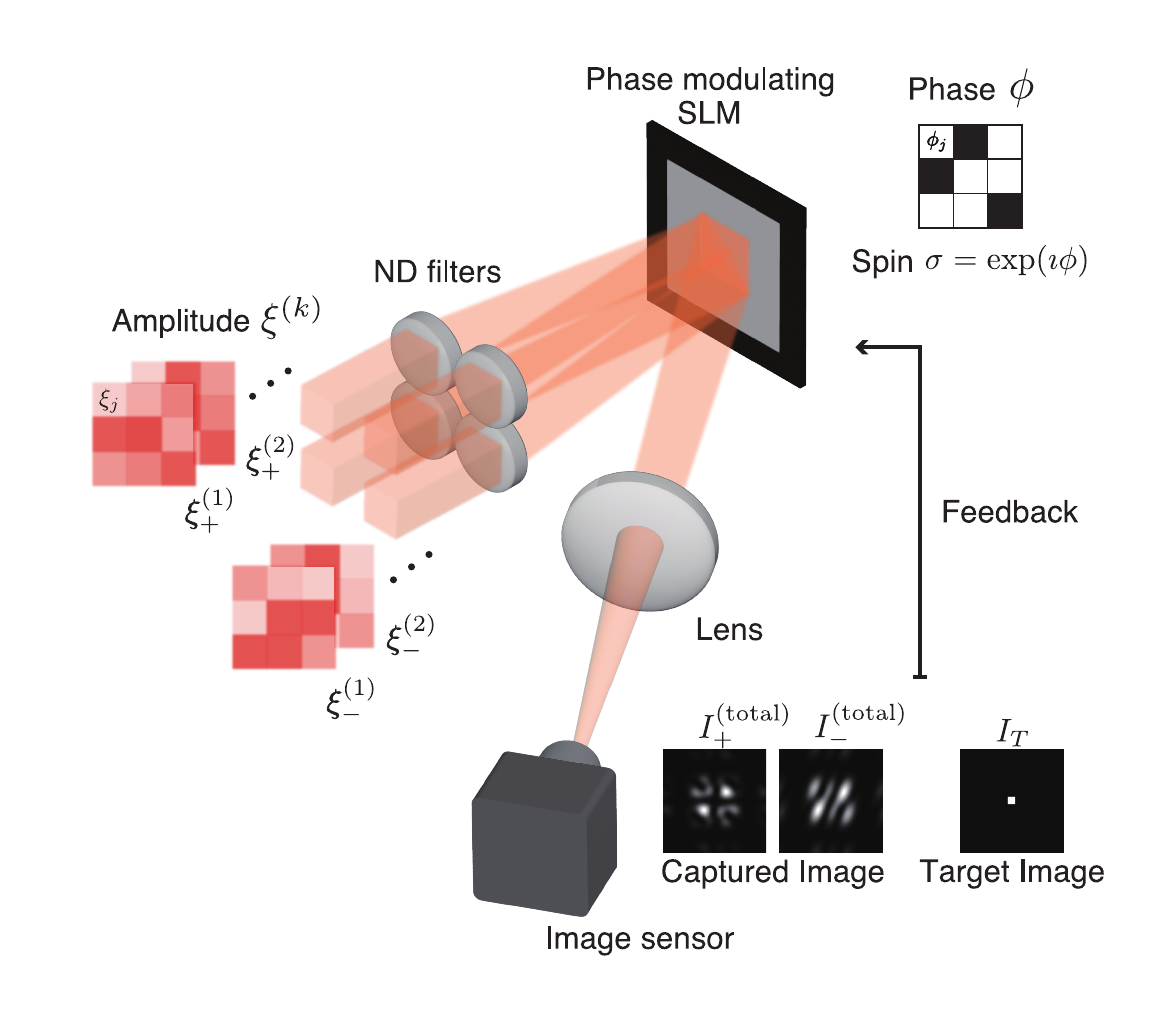}
\caption{Schematic of a spatial-photonic Ising machine by physically tunable space-division multiplexing (SDM-SPIM).}
\label{fig1}
\end{figure}

We consider an implementation of the model of Eq. \eqref{eqHM} with spatial multiplexing; the concept of an SDM-SPIM is depicted in Fig. \ref{fig1}. 
Amplitude distributions $\xi^{(k)}_{j}$, which are mutually incoherent for different $k$, are synthesized by an amplitude modulator with spatial multiplexing. 
The phase-modulated light after the SLM is Fourier transformed and detected as the optical intensity distribution $\bm{I}^{(k)}(x)$, where $x$ is the position of pixels.  
$\bm{I}^{(k)}$ is the optical intensity after being physically tuned with a coefficient parameter $\alpha^{(k)}$ using an optical intensity modulator such as an ND filter. 
The total optical intensity $\bm{I}^{\text{(total)}}$ of the spatial $K$-multiplexed light captured by the image sensor is obtained as a summation $\bm{I}^{\text{(total)}}=\bm{I}^{(1)} +\bm{I}^{(2)}+\cdots+\bm{I}^{(K)}$. 
If the coefficients composing the Ising Hamiltonian $\alpha^{(k)}$ have the same sign, the result can be obtained in a single shot. 
Because the intensity is a non-negative value, if the coefficients contain both positive and negative signs, the intensity should be obtained for each sign. 
In this case, the sum of intensities $\bm{I}^{\text{(total)}}_+$ and $\bm{I}^{\text{(total)}}_-$ of the terms with positive and negative coefficients, respectively, are imaged separately and processed electronically. 
The Ising Hamiltonian can be obtained by capturing the images twice at the maximum. 
Therefore, the computation time in an SDM-SPIM is independent of the rank $K$ of the interaction matrix.

In the primitive SPIM, the Ising Hamiltonian is expressed as $\mathcal{H} \propto\|\bm{I}-\bm{I_T}\|$, where the target intensity $\bm{I_T}$ is a delta function \cite{D.Pierangeli:19}. 
The total optical intensity is normalized by the total power. 
With space-division multiplexing, the normalization is represented as $\bm{I}^{\text{(total)}}_{\text{norm}}=\frac{\bm{I}^{(1)}(x)+ \bm{I}^{(2)}(x)+\cdots+\bm{I}^{(K)}(x)}{\sqrt{\int \left(\bm{I}^{(1)}(x)+ \bm{I}^{(2)}(x)+\cdots+\bm{I}^{(K)}(x)\right)^2\text{d}x}}$.
Then, the Ising Hamiltonian is obtained in the SDM-SPIM as follows.
\begin{equation}
\label{eqHNM}
\mathcal{H}=\left\|\frac{\bm{I}^{(1)}_+(x)+ \bm{I}^{(2)}_+(x)+\cdots+\bm{I}^{(K_+)}_+(x)}{\sqrt{\int \left({\bm{I}^{(1)}_+(x)+\cdots+\bm{I}^{(K_+)}_+(x)}\right)^2\text{d}x}}-\bm{I_T}\right\|-\left\|\frac{\bm{I}^{(1)}_-(x)+ \bm{I}^{(2)}_-(x)+\cdots+\bm{I}^{(K_-)}_-(x)}{\sqrt{\int \left({\bm{I}^{(1)}_-(x)+\cdots+\bm{I}^{(K_-)}_-(x)}\right)^2\text{d}x}}-\bm{I_T}\right\|,
\end{equation}
where $\bm{I}^{(k)}_{+}(x)$ is the intensity of $k$-th term with positive sign.
The value of the Ising Hamiltonian calculated from the optical intensity according to Eq. \eqref{eqHNM} is evaluated to determine the next spin state, the information about which is fed back to the SLM.
The spin state with the ground-state Ising Hamiltonian can be obtained by iterating this procedure.

\subsection{Knapsack Problems}
We used knapsack problems in experiments using the SDM-SPIM. 
The knapsack problems are combinatorial optimization problems to find the subset of items that maximizes the total value by satisfying each knapsack's weight limit. 
This problem cannot be transformed to the Ising model with a rank 1 interaction matrix. 
The 0--1 knapsack problem with integer weights is formulated as follows:
\begin{equation}
\text{maximize} \sum_{i=1}^n v_i x_i,
\end{equation}
\begin{equation}
\text{subject\ to} \sum_{i=1}^n w_i x_i \leq\mathcal{W}, \bm{x}=(x_1, \cdots, x_n)\in {\{0,1\}}^n,
\end{equation}
where $v_i$ and $w_i$ are the value and weight of the $i$-th item for $i = 1, 2, \cdots ,n$ and $\mathcal{W}$ is the knapsack's weight limit.
The corresponding Ising Hamiltonian $\mathcal{H}$ is formulated using a log trick \cite{A.Lucas} as
\begin{align}
\label{eqHS}&\mathcal{H}=\mathcal{H}_A+\mathcal{H}_B,\\
\label{eqHA}&\mathcal{H}_A=A \left(\mathcal{W} - \sum^{n}_{i=1} w_i x_i -  \sum^{m}_{i=1} 2^{i-1} y_i \right)^2,\\
\label{eqHB}&\mathcal{H}_B= -B \sum^{n}_{i=1} v_i x_i,
\end{align}
where $A$ and $B$ are coefficients for the individual terms and $y_i \in \{0,1\}$ are auxiliary variables. 
A conservative choice for the number $m$ of auxiliary variables is $m=\lfloor \log_2\mathcal{W}\rfloor+1$, which can be reduced to $m=\lceil \log_2(\text{max}\ w_i)\rceil$ under the assumption of nontrivial problems where $\sum_i w_i> \mathcal{W}$.
Consistent with the individual terms, Eqs. \eqref{eqHA} and \eqref{eqHB} are the constraint and objective terms. 
The Ising Hamiltonian in Eqs. \eqref{eqHS}-\eqref{eqHB} can be rewritten with the form of size $N=n+m+1$ and rank $K=3$, using the following parameters and variable transformation:
\begin{align}
\label{eqHE}&\mathcal{H}=A\bm{\sigma}^T\bm{\xi}^{(1)}_+\bm{\xi}^{(1)T}_+\bm{\sigma}+ B\bm{\sigma}^T\bm{\xi}^{(2)}_+\bm{\xi}^{(2)T}_+\bm{\sigma}- B\bm{\sigma}^T\bm{\xi}^{(2)}_-\bm{\xi}^{(2)T}_-\bm{\sigma},\\
&\bm{\xi}^{(1)}_+=\frac{1}{2}(w_1, \cdots, w_{N},2^0, \cdots, 2^{m-1},\sum^{n}_{i=1} w_i + \sum^m_{i=1} 2^{i-1} - 2\mathcal{W})^T,\\
&\bm{\xi}^{(2)}_+=\frac{1}{2}(v_1, \cdots, v_{N}, 0, \cdots, 0, 0)^T,\\
&\bm{\xi}^{(2)}_-=\frac{1}{2}(v_1, \cdots, v_{N}, 0, \cdots, 0, 1)^T,\\
&\bm{\sigma} = (2x_1-1,\cdots, 2x_{n}-1,2y_1-1,\cdots, 2y_m-1,1)^T.
\end{align}
Equation \eqref{eqHE} actually contains an additional constant term, which does not essentially change the Ising Hamiltonian.
The coefficients $A$ and $B$ must satisfy the following condition \cite{A.Lucas}, to obtain a solution that adheres to the constraints:
\begin{equation}
0<B\times \text{max}\ v_i <A.
\label{eqratio}
\end{equation}
At the SA process in the experiment, the Metropolis algorithm was used to update the spin states with an updating probability $P(\sigma)=\text{min}\left(1,\exp\left(-\frac{\Delta\mathcal{H}(\sigma)}{T}\right)\right)$ depending on the temperature parameter $T$, where $\Delta\mathcal{H}(\sigma)$ is the difference between the Ising Hamiltonian of the state $\sigma$ and the reference.

\subsection{Optical Setup}
\begin{figure}[t]
\centering\includegraphics[width=11cm]{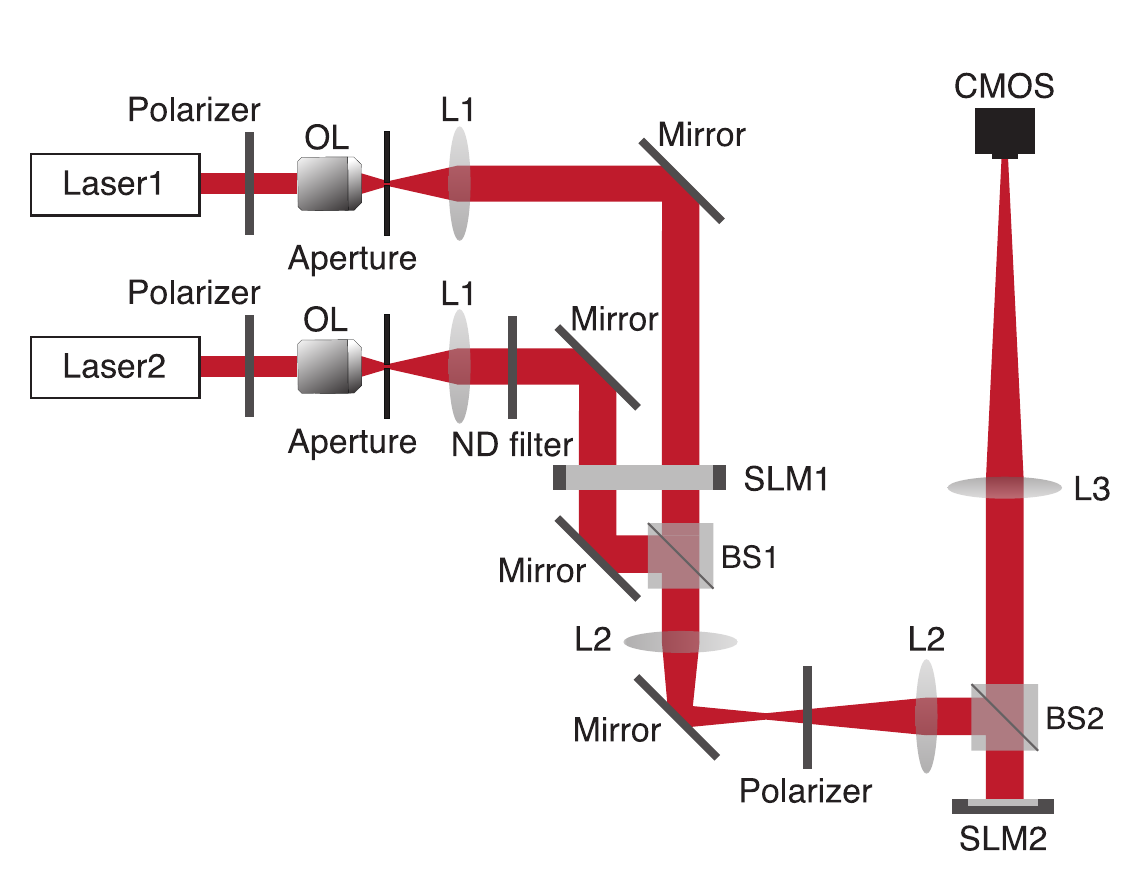}
\caption{Optical setup in the experiment.}
\label{fig2}
\end{figure}

We constructed an SDM-SPIM system and applied it to knapsack problems. 
Figure \ref{fig2} illustrates the optical setup of the two-space-division-multiplexed SPIM.
In this system, two independent optical intensity distributions are required, and thus, two He-Ne lasers (wavelength: 632.8 nm) were used as light sources that are mutually incoherent. 
The individual beams pass through an expander and enter a spatial light modulator SLM1 (Holoeye, LC2012, pixel size $1024\times768$, pixel pitch $36\ \mu \mathrm{m}$).
SLM1 is segmented into two regions to provide an amplitude distribution $\bm{\xi^{(k)}}$ for each beam.
An ND filter is placed in the optical path of beam 2 to adjust the optical intensity ratio between the beams and control the contribution of the constraint and objective terms in the Hamiltonian calculation. 
Without the filter, the intensity ratio of the two terms, i.e., the coefficient ratio $\beta\left(=\frac{B}{A}\right)$, was 4. 
The beams modulated by SLM1 are coaxially superimposed by a mirror and beam splitter (BS1), and their phases are modulated by an SLM2 (Holoeye, Pluto-2, pixel size $1920\times1080$, pixel pitch $8\ \mu \mathrm{m}$). 
The planes at SLM1 and SLM2 have a conjugate relationship given the $4f$ system with two lenses L2 ($f=150\ \mathrm{mm}$). 
Because the size of the corresponding modulation areas between SLM1 and SLM2 must be matched, the minimum size of the modulation area was $360\times360\ \mu \mathrm{m}^2$ ($10\times10$ pixels for SLM1 and $45\times45$ pixels for SLM2). 
The same phase modulation is applied to beams 1 and 2, and the beams are Fourier transformed by lens L3 ($f=300\ \mathrm{mm}$) and captured by a CMOS image sensor (PointGray Research, Grasshopper GS3-U3-32S4, pixel pitch $3.45\ \mu \mathrm{m}$, 16-bit monochrome grayscale). 
The optical intensity $I^{\text{(total)}}$ is obtained by cropping the image to $10\times10$ pixels at the center. 
In the experiment, we used $\left|\xi^{(k)}_{j}\right|$ as the amplitude with flipping the corresponding spin $\sigma_j$ in case that the amplitude modulation $\xi^{(k)}_{j}$ is negative.

\section{Result}
\subsection{Demonstration using knapsack problems}
We investigated the system behavior using a knapsack problem with 13 items (Appendix A: Problem 1) to demonstrate the SDM-SPIM.
In Problem 1, the weight constraint is $\mathcal{W}\leq80$. 
The coefficient ratio was chosen to be $\beta=0.04$ (ND filter transmittance was 1\%) per the requirements of Eq. \eqref{eqratio}. 
The required amplitude corresponding to the fixed spin was 45, but it was divided into three parts of 15 each considering the dynamic range of the modulation amount. 
Consequently, the total number of spins at execution was $N=17$. 
The initial temperature in SA was experimentally determined and the rate of decrease was set to a constant rate.
The solution search was executed with 3000 iterations, and we solved the problem 50 times with randomly determined initial conditions.

\begin{figure}[t]
\centering\includegraphics[width=13cm]{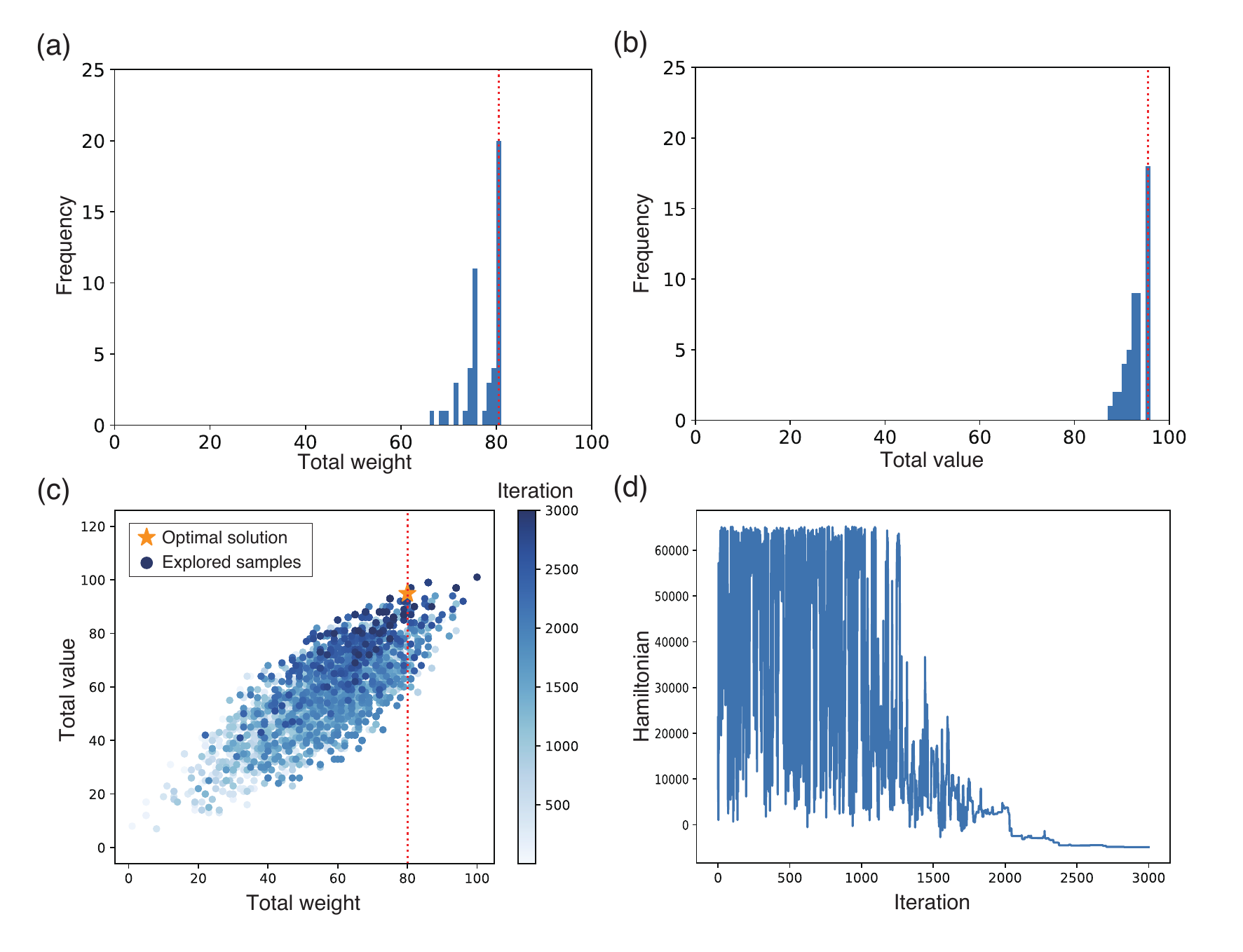}
\caption{Result using the 13-item knapsack problem. (a) Histogram of the total weight for obtained solutions. (b) Histogram of the total value for obtained solutions. (c) Example of the distribution of explored samples. (d) Example of the time evolution of Ising Hamiltonian values. }
\label{fig3}
\end{figure}

Figures \ref{fig3} (a) and \ref{fig3} (b) illustrate histograms of the total weight and total value of the solutions obtained. 
The results demonstrate that the optimal solution (total value: 95) was obtained. 
In the demonstration experiment, the final solution was determined as the sample with the highest value under the constraint weight among all the explored samples.
Figure \ref{fig3} (c) illustrates the sample distributions obtained during iterations. 
The horizontal axis indicates the total weight of each sample, and the vertical axis indicates the total value of each sample.
As the iteration progresses, the search area converges to an area around the optimal solution. 
Figure \ref{fig3} (d) illustrates an example of the time course of the Ising Hamiltonian with iteration. 
As illustrated at the end of the iteration in Figs. \ref{fig3} (c) and \ref{fig3} (d), the near-ground state of Ising Hamiltonian obtained in the experiment certainly corresponds to samples of near-optimal solution.

\subsection{Optical Parameter Tuning}
The coefficient ratio, $\beta=\frac{A}{B}$, of the constraint term to the objective term should be set appropriately, to obtain the optimal solution of a knapsack problem. 
In the SDM-SPIM, this ratio can be controlled by the intensity of the light corresponding to each term. 
We solved a four-item knapsack problem (Problem 2) to investigate the impact on the property of the solution search when changing the transmittance of the ND filter in the optical path of the laser 2. 
The number of the auxiliary variable for Problem 2 was set to $m=4$, and the total spin number was $N=8$ to check that the constraint is still not violated by adding a sizable auxiliary spin.
The temperature parameter in SA was set to decrease at a similar constant rate for all ND filters. 

\begin{figure}[t]
\centering\includegraphics[width=13cm]{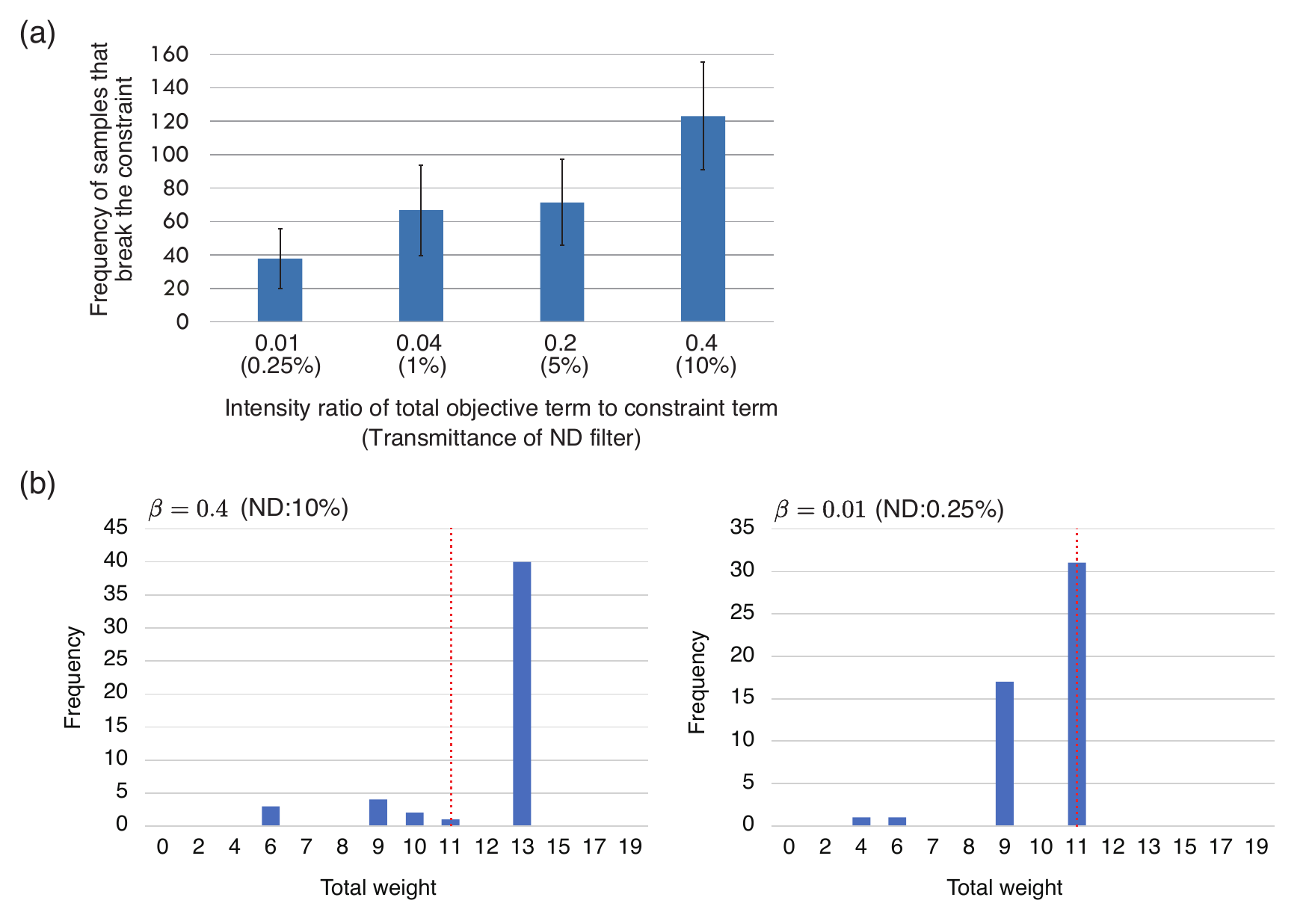}
\caption{Result with ND filter tuning. (a) Frequency of samples that violate the weight constraint in the iterations. (b) Histogram of the total value for obtained solutions with 10\% and 0.25\% ND filters.}
\label{fig4}
\end{figure}

Figure \ref{fig4} (a) illustrates the number of explored samples that violated the weight constraint throughout the iterations, and Fig. \ref{fig4} (b) is the histogram of the total weight for each solution obtained with the ratio $\beta=0.4$ and $0.01$ (ND filter transmittances were 10\% and 0.25\%). 
We solved each problem 50 times. 
The total iteration was 300.
When the optical intensity representing the objective term is decreased, the search is more likely to be run while adhering to the constraint.
In contrast, when the optical intensity is increased, the search is more likely to be run in violation of the constraint. 
In the result, when $\beta=0.01$ (ND: 0.25\%), no solution violated the weight constraint. 
In contrast, when $\beta=0.4$ (ND: 10\%), many solutions were obtained that violated the constraint. 
The experimental result illustrates that the distribution of the obtained solutions changed with the ND filter. 
This indicates that the solution-search space can be manipulated by tuning the optical parameter. 
An effective constraint condition should be $\beta\leq\frac{1}{13}=0.077$, because $\text{max} v_i=13$ in Problem 2, according to Eq. \eqref{eqratio}. 
The result in Fig. \ref{fig4} is consistent with this. 

\begin{figure}[t]
\centering\includegraphics[width=13cm]{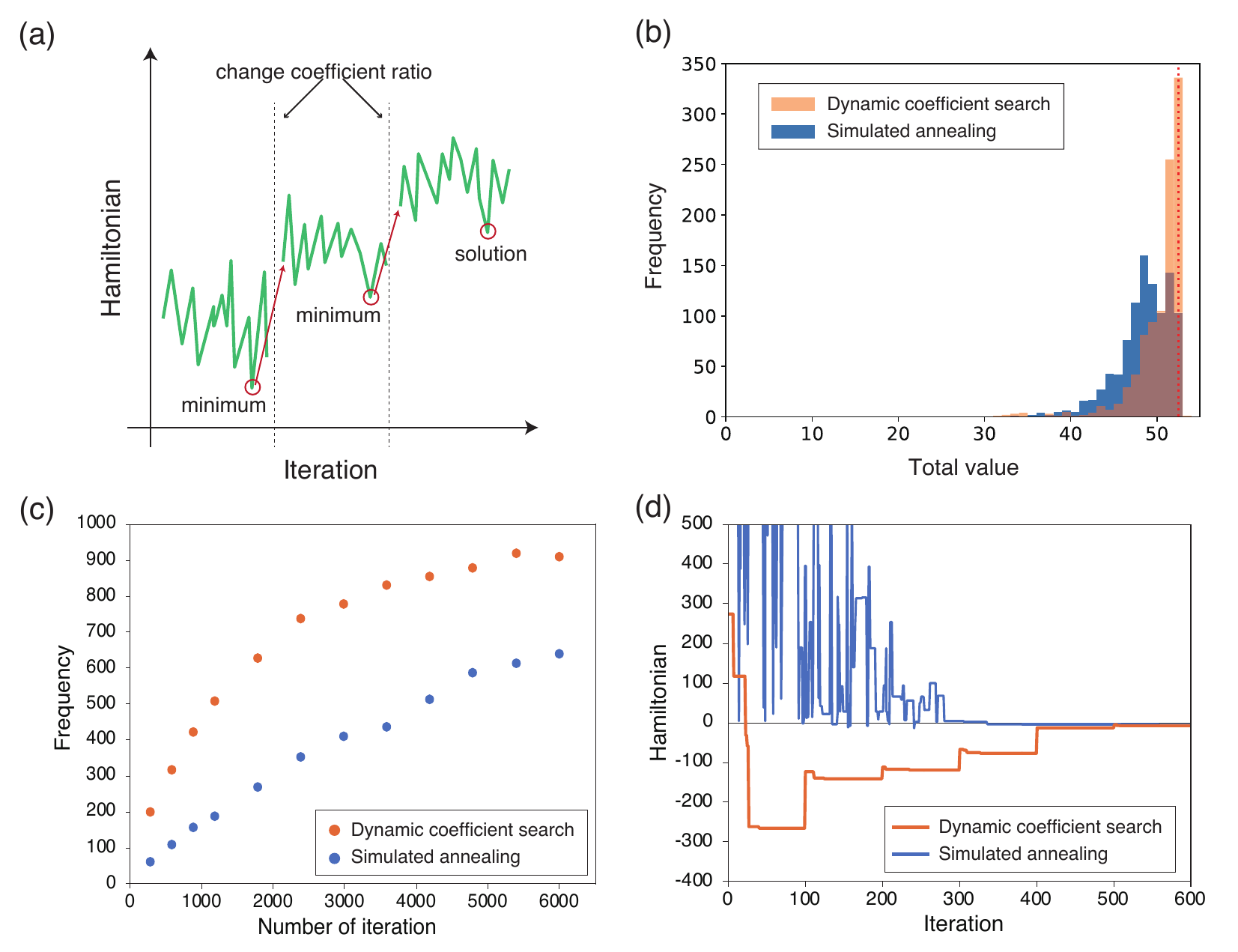}
\caption{Result of numerical simulation of the Dynamic Coefficient Search (DCS). (a) Schematic of the DCS. (b) Histogram of the total value for solutions obtained by the DCS and the SA. (c) Comparison of frequency of the optimal solution obtained between the DCS and the SA. (d) Example of time evolution of Ising Hamiltonian values obtained with the DCS and the SA.}
\label{fig5}
\end{figure}

In Fig. \ref{fig4}, we obtained results using a fixed ND filter transmission during iterations.
However, if the ND filter could be changed for each iteration, it would be possible to change the search trend in steps.
Based on this idea, we investigated a new annealing method, dynamic coefficient search (DCS), where the coefficient ratio is changed step-by-step during iteration. 
Figure \ref{fig5} (a) provides an overview of the DCS process. 
Varying the coefficient ratio in steps during the iteration changes the value of the Ising Hamiltonian for the same spin state. 
The spin state with the minimum Ising Hamiltonian obtained in the search using the same coefficient ratio is adopted as the initial state for the subsequent coefficient ratio.
The spin state is updated by annealing at a constant temperature. 
By repeating this process, the optimal solution is obtained.
Changes in coefficient ratios can be achieved in the optical system by, for example, mechanically changing the ND filters or adjusting a laser with variable emission intensity. 

We compared the SA and the DCS by solving a knapsack problem (Problem 3) with a numerical simulation. 
The coefficient in SA was fixed at $\beta=0.05$. 
The coefficient in the DCS was varied in the order $\beta=2, 1, 0.8, 0.5, 0.1, 0.05$ at each step, and the annealing temperature was kept constant.

Figure \ref{fig5} (b) illustrates the histogram of the total value for solutions obtained in 1000 executions. 
The total number of iterations was set to 600, with 100 iterations for each coefficient in the DCS. 
The results illustrate that DCS is more likely than SA to obtain solutions closer to the optimal solution. 
Figure \ref{fig5} (c) illustrates the frequency of optimal solutions obtained out of 1000 runs when the total number of iterations is varied. 
The trend remained unchanged in this result, with the DCS demonstrating superior properties to the SA. 
A possible reason for the improved result is the difference in the convergence process of the solutions. 
For example, Fig. \ref{fig5} (d) illustrates the time evolution of the Ising Hamiltonian for SA and the DCS. 
In the knapsack problem, the Ising Hamiltonian of the constraint term converges to 0 as the knapsack weight approaches its limit: the smaller the objective term, the higher the value of the selected items. 
Based on Eq. \eqref{eqratio}, a solution with a smaller Ising Hamiltonian may not necessarily be superior because it may violate the constraints.
In the SA with the fixed coefficients, the constraint term is dominant from the initial iterations, and the search is preferentially focused on solutions that satisfy the constraints. 
In contrast, in the DCS, the constraint is weak in the initial iterations, enabling it to search for solutions with a high total value while violating the constraint. 
This property may help to find the optimal solution quickly.

\section{Discussion}
The SDM-SPIM increases the flexibility of the interactions by extending the interaction matrix with spatial resources by acquiring the Ising Hamiltonians in a single or double shot. 
In the demonstration, we applied three amplitude distributions that are spatially and temporally multiplexed to represent a rank-3 interaction matrix. 
At most two time-division processes for individual signs are sufficient. 
Thus, the time required for time-division processing is constant, regardless of the rank of the interaction.

Increasing the degree of spatial multiplexing in the SDM-SPIM required increasing the number of pixels of the amplitude modulator that can be modulated and superimposing the encoded light for the multiplexing number. 
In this study, the optical system was constructed using an SLM, mirrors, and BSs in the multiplexing part. 
However, because of the limitation of the physical size of the elements, the beam illumination and combining part should be made more compact to increase the multiplexing number. 
The optical setup using highly integrated technology will enable the representation of a high-rank problem by relaxing the limitation of encodable spins because of the domain division of an amplitude-modulating SLM.
In contrast, dividing the region of a phase-modulating SLM is unnecessary, so the number of its pixels is not a limiting factor of the multiplexing number.

The DCS considered in this paper seems advantageous not only because of its potential to improve search efficiency but also because it can extend the physically implementable part of the system. 
Part of the DCS process can be physically achieved by adjusting the ratio of the coefficient of each term. 
The SDM-SPIM setup enables the coefficients to be adjusted simply by adjusting the ratio of the intensity of the encoded light.
Hence, it can be introduced without modification into the optical system of SDM-SPIM using variable laser intensities.
Simulated fluctuation in annealing can also be replaced using noise fluctuation during imaging \cite{D.Pierangeli:20a}. 
Other parts of the system, such as the evaluation of the Ising Hamiltonian and control of the devices will still rely on the computer as before.
With the SDM-SPIM setup using intensity-variable lasers and the DCS, the Ising Hamiltonian calculation and electrical portion of the optimization process is physically archivable.

We can also consider other types of problems for which applying the DCS is more effective. 
Although knapsack problems have only one constraint term, the DCS has the potential to select constraints preferentially to be kept in a problem with multiple constraints. 
This potential is expected to be valuable in obtaining approximate solutions sufficient for practical use.

\section{Conclusion}
We presented a SPIM system with physically tunable space-division multiplexing as one of the implementing methods to physically calculate the weighted sum of the Ising Hamiltonian components at a time in a multi-component computing model.
We experimentally demonstrated the validity of the SDM-SPIM by solving knapsack problems with a constraint term using the system. 
Another result demonstrates that the solution distribution obtained by the SDM-SPIM can be changed by controlling the intensity ratio of the encoded light using an ND filter placed in the optical path. 
Based on this property, we considered a new DCS algorithm as a solution-search method and compared it with the SA by simulation. 
The results suggest the ability to improve the performance of the SDM-SPIM by dynamically tuning optical parameters.
Our method computes the Ising Hamiltonian in a single or double shot, saving computation time for high-rank problems.
The SDM-SPIM that can handle a large number of spins will enable the solving of various combinatorial optimization problems at high speed by increasing the number of multiplexes.

\bigskip
\section*{Appendix A: Detail of knapsack problems}
A knapsack problem generator \cite{D.Pisinger} was used to obtain three problem instances for the experiment.

Problem1:
\begin{align*}
&n = 13, \mathcal{W} = 80,\\
&\bm{v} = (7, 7, 8, 8, 2, 7, 12, 4, 9, 14, 2, 7, 14),\\
&\bm{w} = (6, 7, 1, 15, 14, 8, 5, 6, 4, 7, 5, 12, 10).
\end{align*}
The optimal solution to this problem is $\bm{x}=(1, 1, 1, 1, 0, 1, 1, 0, 1, 1, 1, 1, 1)$, for which the total weight is 80 and the total value is 95. 
The number of auxiliary variables is $m = 4$ using the log trick.

Problem2:
\begin{align*}
&n = 4, \mathcal{W} = 11,\\
&\bm{v} = (6, 10, 12, 13),\\
&\bm{w} = (2, 4, 6, 7).
\end{align*}
The optimal solution to this problem is $\bm{x}=(0, 1, 0, 1)$, for which the total weight is 11 and the total value is 23. 
The number of auxiliary variables is $m = 4$ using the log trick. 

Problem3:
\begin{align*}
&n = 10, \mathcal{W} = 60,\\
&\bm{v} = (20, 18, 17, 15, 15, 10, 5, 3, 1, 1),\\
&\bm{w} = (30, 25, 20, 18, 17, 11, 5, 2, 1, 1).
\end{align*}
The optimal solution to this problem is $\bm{x}=(0, 0, 1, 1, 1, 0, 1, 0, 0, 0)$, $(0, 0, 1, 1, 1, 0, 0, 1, 1, 1)$, $(0, 0, 1, 1, 0, 1, 1, 1, 1, 1)$, $(0, 0, 1, 0, 1, 1, 1, 1, 1, 1)$, for which the total weight is 57--60 and the total value is 52. 
The number of auxiliary variables is $m = 5$ using the log trick.

\bigskip
\begin{backmatter}

\bmsection{Funding}
Japan Science and Technology Agency (JPMJCR18K2); Japan Society for the Promotion of Science (JP23H04805).

\bmsection{Disclosures}
The authors declare no conflicts of interest.

\bmsection{Data availability} 
Data underlying the results presented in this paper are not publicly available at this time but may be obtained from the authors upon reasonable request.
\end{backmatter}


\bibliography{MultiplexedSPIM}

\end{document}